\documentclass[aps,prc,showpacs,twocolumn,preprintnumbers,floatfix, showkeys,tightenlines]{revtex4}
\usepackage{amsmath,amssymb,amsfonts}
\usepackage{hyperref}
\usepackage{graphicx}
\usepackage{color}

\allowdisplaybreaks
\begin{document}

%\title{Anatomy of the density content of the nuclear  symmetry energy and incompressibility }
\title{Equation of state of nuclear matter from empirical constraints }

\author{N. Alam}
\address{Saha Institute of Nuclear Physics, 1/AF Bidhannagar, Kolkata
{\sl 700064}, India}
\author{B. K. Agrawal}
\address{Saha Institute of Nuclear Physics, 1/AF Bidhannagar, Kolkata
{\sl 700064}, India}
\author{J. N. De}
\address{Saha Institute of Nuclear Physics, 1/AF Bidhannagar, Kolkata
{\sl 700064}, India}
\author{S. K. Samaddar}
\address{Saha Institute of Nuclear Physics, 1/AF Bidhannagar, Kolkata
{\sl 700064}, India}
\author{G. Col\`o}
\address{ Dipartimento di Fisica, Universit\`a
degli Studi di Milano, via Celoria 16, I-20133 Milano, Italy}
\address{INFN, sezione di Milano, via Celoria 16, I-20133 Milano, Italy}

\begin{abstract}

From empirically determined values of some of the characteristic
constants associated with homogeneous nuclear matter at  saturation
and sub-saturation densities, within the framework of a Skyrme-inspired
energy density functional, we construct an equation of state (EoS) of
nuclear matter.This EoS is then used to predict values of density slope
parameters of symmetry energy $L(\rho)$, isoscalar incompressibility
$K(\rho )$ and a few related quantities.  The close consonance of
our predicted values with the currently available ones for the density
dependence of symmetry energy and incompressibility gleaned from diverse
approaches offers the possibility that our method may help in settling
their values in tighter bounds.  Extrapolation of our  EoS at supranormal
densities shows that it is in good harmony with the one extracted
from  experimental data.

\end{abstract}
\maketitle

%\texttt{Mean-field model, Nuclear matter, Symmetry energy, Equation of state, Neutron stars}

\section{Introduction}

The nuclear equation of state (EoS) entails in a broad sweep knowledge of
the diverse properties of nuclear matter: its saturation density, energy
per nucleon  $e(\rho )$ ($\rho $ is the density), incompressibility
$K(\rho )$, the symmetry energy and its density content, $\it
i.e.,$ the symmetry coefficient $e_{sym}(\rho )$, the symmetry slope
parameter $L(\rho )$, the symmetry incompressibility $K_{\tau }(\rho )$
and all the higher symmetry derivatives. A cultivated attention is
naturally drawn in recent times to  have a refined understanding of
this nuclear EoS from both experimental and theoretical sides. The
binding energies of stable atomic nuclei are the most accurately
known experimental entities in nuclear physics, these supplemented
with knowledge of giant monopole and dipole resonances followed
by theoretical analysis have yielded some of the EoS parameters
like the saturation density $\rho_0$ of symmetric nuclear matter and
$e(\rho_0), K(\rho_0) $ and $e_{sym}(\rho_0)$ in reasonably tight bounds
\cite{Moller12,Myers69,Myers80,Jiang12,Fan14,Liu10}.  The knowledge
of the symmetry derivatives $L(\rho_0)$ and $K_{\tau }(\rho_0)$ is
still not very certain \cite{Moller12,Trippa08,Carbone10,Dong12}.
Analyses of the different nuclear observables do not help much  in
removing the uncertainty. Correlation systematics of nuclear isospin
with neutron skin thickness \cite{Centelles09,Warda09}, isospin
diffusion \cite{Chen05,Li08}, nucleon emission ratios \cite{Famiano06}
or isoscaling \cite{Shetty07} in heavy ion collisions, all yield values
of the symmetry slope parameter $L_0 (=L(\rho_0)) $ that are not much
in consonance with one another.  Some attempt was recently made to
constrain $L_0$ in tighter bounds from nuclear masses aided by microscopic
calculations \cite{Agrawal12,Agrawal13} on neutron skin of heavy nuclei,
it was found to give $L_0$ =59 $\pm $ 13.0 MeV.  Other recent attempts,
from analysis of the isovector giant dipole and quadrupole resonances in
$^{208}$Pb nucleus \cite{Roca-Maza13a,Roca-Maza13} give $L_0$ = 43$\pm26$
and  37$\pm $18 MeV respectively. This underscores the still unresolved
uncertainty in getting to the value of $L_0$ and asks for newer avenues
to understand it.  The present state of the art on symmetry energy and
related parameters can be found in the topical issue on nuclear symmetry
energy \cite{EPJA14}.

The EoS parameters so mentioned pertain to only one density, the
saturation density $\rho_0$. If all of them are known precisely,
it is in principle possible to construct with a suitable energy
density functional (EDF) the nuclear EoS $e(\rho,\delta )$ where $\delta
=(\rho_n-\rho_p)/(\rho_n+\rho_p)$ is the isospin asymmetry. Whereas the
high density end of the EoS would be of immediate value in understanding
the dynamical evolution of the core collapse of a massive star and
the associated explosive nucleosynthesis \cite{Steiner05,Janka07} or
the radii and lower bound of the  maximum mass of cold neutron stars
\cite{Roberts12}, the low-density end helps in getting a closer estimate
of the neutron skin thickness or the neutron density distribution
\cite{Centelles09,Warda09,Agrawal12,Agrawal13} in neutron-rich
nuclei. Further information on the EoS parameters at densities other than
$\rho_0$ may  put the nuclear EoS on firmer grounds, unfortunately, there
are only a few of them. At density higher than $\rho_0$, information
from experimental data has still large uncertainty \cite{Xiao14};
at sub-saturation density, from giant dipole resonance analysis, a
quantitative constraint on $e_{sym}(\rho )$ could be put as 23.3 MeV
$< e_{sym}(\rho =0.1 \,{\rm fm}^{-3}) <$24.9 MeV \cite{Trippa08}.  Further
information derived from theoretical analyses at around this density may
be of added significance, (i) the energy per nucleon of neutron matter
is $\sim  10.9\pm0.5$ MeV at $\rho $= 0.1 fm$^{-3}$ \cite{Brown13} and
(ii) the density derivative of the nuclear incompressibility $M_c=3\rho
dK/d\rho |_{\rho=\rho_c}  \simeq $ 1100$\pm $70 MeV where $\rho_c$
is $\simeq $ 0.7$\rho_0$ \cite{Khan12}.

As is evident from the previous discussion, the plethora of nuclear
EoS failed to effectively well-constrain the density content
of the nuclear symmetry energy and the nuclear incompressibility
from fits to diverse microscopic nuclear data. The reason lies
in the choice of different sets of microscopic observables to
be fitted. The isoscalar and isovector quantities associated with
nuclear matter, however, have emerged to be very well-constrained.
 The isoscalar
entities are i) $e_0(=e(\rho_0)=-16.0\pm $0.1 MeV), ii) saturation density
$\rho_0 $(=0.155 $\pm $0.008 fm$^{-3}$), where the pressure $P(\rho_0)=0$,
iii) the incompressibility coefficient $K_0(=K(\rho_0)=9\partial^2
e/\partial \rho^2|_{\rho_0} = 9dP/d\rho|_{\rho_0}$ =240 $\pm $20
MeV) \cite{Shlomo06}. All these quantities refer to symmetric nuclear
matter. The isovector quantities are iv) $e_{sym}(\rho_0)$(=32.1$\pm
$0.31 MeV)\cite{Jiang12} and v) $e_{sym}( \rho =0.1$ fm$^{-3})$(=24.1
$\pm $0.8 MeV)\cite{Trippa08}.   

In this article, we have tried to find how the input of the empirical
knowledge of these quantities can be used to construct an EDF for
nuclear matter and to predict the as yet not so well-constrained density
dependence of its symmetry properties in reasonably tighter bounds. These
isoscalar and isovector nuclear parameters effectively contain condensed
experimental information on the bulk nuclear properties.  As opposed
to direct investigation of the microscopic properties of nuclei as
done, {\it e.g,} in Ref. \cite{Agrawal05} which can lead to somewhat
different predictions depending on the observables chosen to be explored,
the alternate edifice for the nuclear EDF built in this article on the
well established nuclear bulk parameters is so structured, as we see
later, that it gives an easy and transparent look to the correlations
of the predicted values of the density derivatives of the symmetry
energy and the nuclear incompressibility to the input parameters. Furthermore, if
the values of the density derivatives $L_0, K_\tau , M_0 $
etc could be well settled by as yet other unexplored means and differ
from our predicted values, the foundation {\it i.e.,} the values of
the nuclear bulk parameters or the choice of the Skyrme EDF then become
subject of fresher scrutiny. Computationally, our method is also much
less intensive.  To our knowledge, a comprehensive study of this kind
has not been done before.

Henceforth, the quantities corresponding to the densities $\rho_0$ and
$\rho=$0.1 fm$^{-3}$ would be denoted with the subscripts '0' and
'1', respectively (like $e_{sym,0}$, $e_{sym,1}$ etc.). 
The value of $\rho_0$ is an indirectly
obtained entity. From acceptable Skyrme energy density functionals,
it is $\sim$ 0.16 fm$^{-3}$ \cite{Dutra12} where as the relativistic mean-field models
give a value of ${\rho_0}$ in the vicinity of $\sim$ 0.15 fm$^{-3}$
\cite{Serot86,Serot97}. Our choice for $\rho_0$ covers this range.

The paper is organized as follows. Sec. II contains a brief discussion of
the theoretical elements. Results and discussions are presented in Sec.
III. Sec. IV contains the concluding remarks.

\section{Theoretical edifice}

The Skyrme framework is chosen for the
energy density functional 
\cite{Brack85}. The energy per nucleon for nuclear matter is then,
 \begin{eqnarray}
e(\rho,\delta)& =& a_1\left [\left(\frac {1+\delta }{2}\right)^{5/3}+\left(\frac {1-\delta
}{2}\right)
^{5/3}\right ]\rho^{2/3} \nonumber \\
&&+(b_1+b_2\delta^2)\rho +(c_1+c_2\delta ^2)\rho^{{\alpha+1}} \nonumber \\
&&+\left[d_1 \left \{ \left(\frac {1+\delta}{2}\right)^{5/3}+ \left(\frac {1-\delta
}{2}\right)^{5/3}\right  \} 
\right . \nonumber \\
&& \left . +d_2 \left \{\left (\frac {1+\delta }{2}\right)^{8/3}+ \left(\frac {1-\delta }{2}\right)^{8/3}
\right \}\right]
\rho^{5/3}. 
\end{eqnarray}
The first term on the right hand side is the free Fermi gas energy;
$a_1= \frac {\hbar^2}{2m}\frac {3}{5} (3\pi^2)^{2/3}$ =119.14 MeV
fm$^2$, where $m$ is the nucleon mass.  For the chosen values of $e_0$
and $\rho_0$, values of $\alpha$ ranging only from 1/6 to 1/3  allow
for an acceptable set of ($m^*/m$, $K_0$) \cite{Cochet04a} where $m^*$
is the nucleon effective mass. We therefore chose $\alpha = 0.2 \pm 0.1$,
this allows $m^*/m$ to lie in the acceptable range $m^*/m \simeq 0.8\pm
0.2$ \cite{Dutra12}. We take the median value of $\alpha =$0.2.
We are left with six unknown parameters, namely, $b_1, c_1, d_1, b_2, c_2$ and
$d_2$ which would completely define the EDF. As already mentioned,
we have, however, five equations,  three from 
isoscalar entities and two from  isovector 
entities. 

The isoscalar equations are
\begin{eqnarray}
e_0=\frac{a_1}{2^{2/3}}\rho_0^{2/3}+b_1\rho_0+c_1\rho_0^{\alpha +1}+(
\frac{1}{2^{2/3}}d_1+\frac{1}{2^{5/3}}d_2)\rho_0^{5/3},
\end{eqnarray}
\begin{eqnarray}
P_0=0&=&\rho_0^2 [\frac{2}{3}\frac{a_1}{2^{2/3}}\rho_0^{-1/3}+b_1+
c_1(\alpha +1 )\rho_0^\alpha \nonumber \\
&&+\frac{5}{3}(\frac{1}{2^{2/3}}d_1+
\frac{1}{2^{5/3}}d_2)\rho_0^{2/3}],
\end{eqnarray}
and
\begin{eqnarray}
K_0&=&9[\frac{10}{9}\frac{a_1}{2^{2/3}}\rho_0^{2/3}+2b_1\rho_0+
(\alpha +1)(\alpha +2)c_1\rho_0^{\alpha +1} \nonumber \\
&&+\frac{40}{9}(\frac{1}{2^{2/3}}d_1+\frac{1}{2^{5/3}}d_2)\rho_0^{5/3}].
\end{eqnarray}
The isoscalar equations yield  the values of $b_1,c_1$ 
and $(\frac {d_1}{2^{2/3}}+\frac {d_2}{2^{5/3}})$. 
The isovector equations, evaluated at $\rho_0$ and $\rho_1$ (=0.1 fm$^{-3}$)
are given by
\begin{eqnarray}
e_{sym}(\rho )&=&\left .\frac {1}{2} \frac {\partial^2e(\rho )}
{\partial \delta^2}\right|_ {\delta =0} \nonumber \\
&&=\frac {5}{9}\frac {a_1}{2^{2/3}}\rho^{2/3}+b_2\rho +c_2\rho
^{\alpha +1}   \nonumber \\
&&+\left [\frac {5}{9} \frac {d_1}{2^{2/3}}+\frac {20}{9} \frac {d_2}{2^{5/3}} \right]
\rho^{5/3}.
\end{eqnarray}
To fix the remaining parameters, we need an extra condition in conjunction
with the two isovector equations. For this, we take that the energy
per particle of isospin asymmetric nuclear matter is quadratic in
the asymmetry parameter $\delta $.  This condition is found to be an
excellent approximation from nearly all energy density functionals
\cite{Chen09,Constantinou14} and also from microscopic calculations
in the Bruckner-Hartree-Fock (BHF) formalism \cite{Lee98,Vidana02} at
all densities up to $\rho_0$ and a little beyond.  This implies that the
difference between the symmetry energy coefficients defined by Eq.~(5)
and the one by following equation
\begin{eqnarray}
\tilde e_{sym}(\rho )&=&e(\rho,\delta =1)-e(\rho, \delta =0). \nonumber \\
&=& a_1\left (1-\frac {1}{2^{2/3}} \right)\rho^{2/3}+b_2\rho 
+c_2\rho^{\alpha +1}  \nonumber \\
&&+ \left [d_1\left (1-\frac {1}{2^{2/3}}\right)+d_2\left (1
-\frac {1}{2^{5/3}}\right) \right]\rho^{5/3},
\end{eqnarray}
should be minimal.

To achieve this minimality, we take the help of the equation for the symmetry
slope parameter $L$. From its definition
$L(\rho )=3\rho \partial e_{sym}/\partial \rho $, one gets from Eq.~(5),
\begin{eqnarray}
L(\rho)&=&3\rho \left [\frac {10}{27}\frac {a_1}{2^{2/3}}\rho^{-1/3}
+b_2+(\alpha +1)c_2\rho^\alpha \right.  \nonumber \\
&&\left .+\frac {5}{3}\rho^{2/3}\left (\frac {5}{9} \frac {d_1}{2^{2/3}}
+\frac {20}{9} \frac {d_2}{2^{5/3}}\right)\right],
\label{eq:L1}
\end{eqnarray}
With a given value of $L_0$ (=$L(\rho_0)$), from known
$e_{sym,0}$ and $e_{sym,1}$, at the two densities
$\rho_0$ and $\rho_1$, one can solve for $b_2,c_2$ and
$[\frac{5}{9}\frac{d_1}{2^{2/3}}+\frac{20}{9}\frac{d_2}{2^{5/3}}]$.
Since $(\frac{d_1}{2^{2/3}}+\frac{d_2}{ 2^{5/3}})$ is known from the
isoscalar equations, $d_1$ and $d_2$ are now obtained.
\begin{table}[t]
\caption{\label{tab1}Parameters of the energy density functional corresponding to the
central values of the isoscalar and isovector inputs.}
\begin{tabular}{|c|c|c|c|}
\hline
$a_1(\, {\rm MeV fm}^{2})$&119.14&$\alpha$&0.2\\
$b_1(\, {\rm MeV fm}^{3})$&-816.95& $b_2(\, {\rm MeV fm}^3)$&744.65\\
$c_1(\, {\rm MeV fm}^{3(\alpha+1)})$&724.51& $c_2(\, {\rm MeV
fm}^{3(\alpha+1)})$&-1149.66\\
$d_1(\, {\rm MeV fm}^{5})$&-32.99& $d_2(\, {\rm MeV fm}^{5})$&891.15\\
\hline
\end{tabular}
\end{table}

\begin{figure}
\includegraphics[width=0.85\columnwidth,angle=0,clip=true]{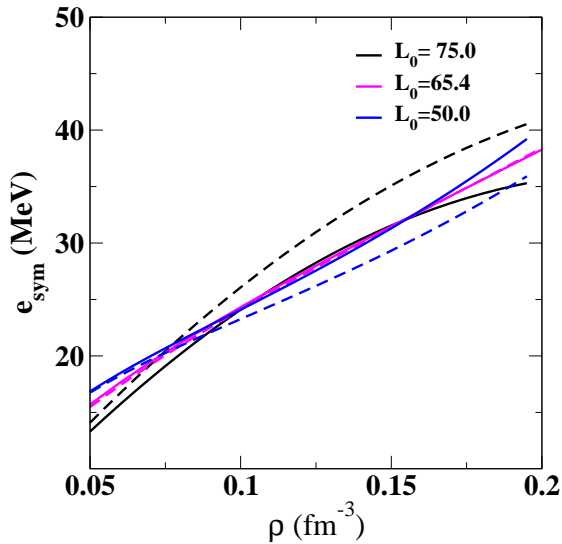}
\caption{(Color online)
{\label{fig1} Plots for the variation of $e_{\rm sym}$ as a function
of density $\rho$. The values of $e_{\rm sym}$ represented by full and
the dashed lines are obtained using two different definitions as given
by Eqs. (5) and (6).  The blue, magenta and black lines corresponds to
$L_0 = 50, 65.4$ and 75 MeV at $\rho = \rho_0$. } With $L_0 = 65.4$ MeV,
one  magenta line falls over the other, they can not be distinguished
from each other.} \end{figure}

Now that for a given $L_0$ all the parameters of the EDF are known,
one can calculate $e_{sym} (\rho )$.  With the same set of coefficients
$\tilde e_{sym}(\rho )$ can also be calculated.  Normally, for an
arbitrary value of the  given $L_0$, $e_{sym}$ and $\tilde e_{sym}$ may
not be equal, only for a specific value of $L_0$, they tend to be equal
(see Fig. \ref{fig1}). To be more specific, as a function of input $L_0$,
we have calculated the coefficients $ b_1, b_2 $ etc. using the relevant
equation for $e_{sym}$ for a set of densities $\rho_i$ lying in the range
0.05 fm$^{-3} < \rho <$ 0.2 fm$^{-3}$, calculated $\tilde e_{sym}$ with
the same set of coefficients and have chosen that $L_0$ as the requisite
one that gives the minimum of $\sum_i [e_{sym}( \rho_i)-\tilde e_{sym}
(\rho_i)]^2 $.  This settles the EoS.  In Fig. \ref{fig1}, $e_{sym}(\rho
)$ and $\tilde e_{sym}(\rho )$ are displayed as a function of
input $L_0$. The full lines refer to $e_{sym}(\rho)$, the dashed lines
to $\tilde e_{sym}(\rho )$.  The difference between these two is
minimum when $L_0$ =65.4 MeV.  All the parameters $a_1, b_1, b_2 $ etc,
corresponding to the density functional are listed in Table \ref{tab1}.

\section{Results and Discussions}

Once the parameters of the EDF (Eq.~1) are known, the higher order
density derivatives of energy and symmetry energy of nuclear matter
may be obtained. They are presented in subsection {\bf A}. This EDF
can also be used to estimate certain properties of microscopic nuclei
like neutron skin thickness of heavy nuclei. This is  discussed in
subsection {\bf B}. As an aside, the EDF has also been employed to
explore some properties of neutron stars, discussion of
which is contained  in subsection
{\bf C}.  

\subsection{Nuclear matter : density derivatives of symmetry  energy
and isoscalar incompressibility}

Expressions for higher order symmetry derivatives $K_{sym,0}$ and $K_{\tau}$
are given by
$K_{sym,0}= 9\rho_0^2\frac {\partial
^2e_{sym}}{\partial \rho^2} |_{\rho_0}$ 
and $K_{\tau }= 9\rho^2_{\delta
}\frac {\partial ^2e_{sym}} {\partial \rho^2} |_{\rho_{\delta }}$.  Here
$\rho_{\delta }$ is the saturation density of asymmetric nuclear matter
corresponding to the asymmetry $\delta $. The symmetry derivative
$K_{sym,0}$ and $K_{\tau }$ are related: $K_{\tau }=K_{sym,0}-6 L_0-\frac
{Q_0L_0}{K_0}$, where $Q_0=27\rho_0^3\frac {\partial
^3e_(\rho, 0)}{\partial \rho^3} |_{\rho_0}$. 
They all can be evaluated from the EDF parameters.
The density derivative of the
isoscalar incompressibility $M(\rho )(=3\rho \frac {dK(\rho )}{d\rho })$
of symmetric nuclear matter is also calculated. At the saturation density
$\rho_0$, $M_0 (=M(\rho_0 ))$ equals $12K_0+Q_0$. 
In Table \ref{tab2} we list the calculated values  of various isoscalar and
isovector quantities together with their total uncertainties. The latter
are  associated with  the uncertainties in the six input quantities
$Y_i (=\{e_0,\rho_0, K_0,e_{\rm sym,0}, e_{\rm sym,1}
 $ and $\alpha\})$.  The extracted value of $L_0$ is
65.4 $\pm $13.5 MeV. It is in excellent consonance with that obtained
from analysis of pygmy dipole resonance \cite{Colo14} and in very good
agreement with that obtained earlier from nuclear masses and the neutron
skin thickness of heavy nuclei \cite{Agrawal12,Agrawal13}. This is also
very consistent with the value $L_0$=66.5 MeV obtained from a systematic
analysis within the BHF approach using a realistic
nucleon-nucleon potential \cite{Vidana09}. 
 Not much can be said about the reasons behind the good agreement
between our present extracted value of $L_0$ with that obtained from
pigmy dipole resonance \cite{Carbone10} except that in this case
the value of $e_{sym}(\rho_0)$ matches extremely well with our
input value. The agreement with that obtained from BHF approach
\cite{Vidana09} 
is possibly coincidental; the one aspect that is to be noted here
is that all the symmetry derivatives $L_0, K_{sym,0}$ and $K_\tau $
from the BHF approach are in extremely good consonance with our 
calculated values though the nuclear bulk parameters ($\rho_0, e_0,
K_0 $ and $e_{sym}(\rho_0 )$) do not have a good common overlap.
Most of the uncertainty in our
extracted value of $L_0$ comes from the uncertainties in the empirically
obtained quantities $\rho_0$ and $e_{sym}(\rho )$ at the two densities
(see also Table \ref{tab3}).  If the central value of $\rho_0$ is pushed
down to 0.14 fm$^{-3}$ keeping all other input parameters same, then the
central value of $L_0$ shoots up to 86.1 MeV.  Attention is also drawn
to the calculated value of $K_{\tau }$. Analyzing the experimental
breathing-mode energies of Sn-isotopes, Li $\it et. al., $\cite{Li07}
suggested its value as  $-$550 $\pm $100 MeV. This is pointed to be
too strongly negative to be compatible with the behavior of low-density
neutron matter \cite{Piekarewicz09,Piekarewicz10}.  Higher order effects
such as surface symmetry, present in such analysis in disguise, may have
contributed to such a high value. Explicit inclusion of  the surface
symmetry term seems to lower the value of $K_\tau $ to $\sim -$350
MeV \cite{Pearson10,Majumdar94}.  Our present value of $K_\tau $=
$-$321.6 $\pm $34.4 MeV is in very good agreement with this; it is also
in close consonance with the value of $-$370 $\pm $120 MeV extracted from
measurements of isospin diffusion in heavy ion collisions. The empirical
value of $M_c (= 1100 \pm 70\,{\rm MeV})$ obtained from analysis of
Giant Monopole Resonance Energies of Sn-isotopes and of $^{90}$Zr and
$^{144}$Sm nuclei \cite{Khan12} is very compatible with our calculated
value; similarly, the value of $Q_0$ calculated by us conforms well with
the one ($Q_0 = -350 \pm 30 \,{\rm MeV}$) obtained from examination of
a host of standard Skyrme interactions \cite{Chen09}.

The total uncertainties in the various quantities considered in Table \ref{tab2} are
evaluated  as \cite{Arfken05},
\begin{equation}
\Delta X = \sqrt{\sum_i \left (\Delta X_i\right)^2}.
\end{equation}
where, $\Delta X_i = \frac{\partial X}{\partial Y_i}\Delta Y_i$;
$\Delta X$ is the total uncertainty on a given  quantity $X$
induced by the associated uncertainties  $\Delta Y_i(= 0.1 {\rm MeV},
0.008 {\rm fm}^{-3}, $20 {\rm MeV}, 0.31 {\rm MeV}, 0.8 {\rm MeV}$,
0.1)$ in the input quantities $Y_i$.
The quantities $\frac{\partial X}{\partial Y_i}$ are calculated
numerically, their signs reflect the direction of change in $X$ with
increase in $Y_i$.  
Table \ref{tab3} displays, for the relevant observables $X$, the values
of $\frac{\partial X}{\partial Y_i}$ along with the associated total uncertainty
$\Delta X$.
The derivatives $\frac{\partial X}{\partial Y_i}$
help in estimating the partial contributions $\Delta X_i $ to the
total uncertainty $\Delta X$. Once $\frac{\partial X}{\partial Y_i}$
are known, it is easy to estimate the change in $\Delta X$ with change
in $\Delta Y_i$.  This table can be an instructive guide in improving
existing energy density functionals as it enables one to understand how
the various quantities $X$s can be adjusted by changing $Y_i$s or vice
versa.  One may note that the symmetry observables $L_0, K_{\rm
sym,0}$ and $K_\tau$ correlate with $e_{\rm sym,0}$ and $e_{\rm
sym,1}$ always in the reverse (see columns 6 and 7 in table \ref{tab3}).
This correlated structure of $L_0$ on $e_{\rm sym,0}$ and $e_{\rm sym,1}$
has been noticed earlier \cite{Ducoin11}. Similar correlation of the
symmetry observables  $K_{\rm sym,0}$ and $K_\tau $ on $e_{sym,0}$
and $e_{sym,1}$ is noticed in our calculation.

The fractional  contributions $x_i^2 = \left (\frac{\Delta X_i}{\Delta
X}\right )^2$ 
=$\left ( \frac{\partial X}{\partial Y_i} \frac{\Delta Y_i}{\Delta X}
\right )^2 $
to the uncertainties in the
observables $X$ from the uncertainties in the input quantities $Y_i$ are
shown in Fig.~2 in color code.
As one sees, $\sum_i x_i^2 = 1$. They  depict
the relative importance of the precision of the input parameters in
measuring up the uncertainties in an observable $X$. From the first column 
in the figure, on can easily see that the  
uncertainty in energy per particle $e_0$ has a negligible role in the
uncertainties in the observables $X$ we calculate. One also sees that 
nearly all the uncertainties in $M_0$ 
and $M_c$ emanate from the uncertainty in
$K_0$  and that the uncertainty 
in $\alpha$ has a very strong role in the evaluated
uncertainty of $K_\tau $. 

\begin{table}[t]
\caption{\label{tab2}Values of the extracted entities from the nuclear
EoS.  All quantities  are in MeV.}
\begin{tabular}{|c|c|c|c|}
\hline
$L_0$&$65.4\pm13.5$&$M_c$&$1150\pm91$\\
$K_{sym,0}$ &$-22.9\pm73.2$& $Q_0$&$-344\pm56$\\
$K_{\tau }$ &$-321.6\pm34.4$& $M_0$  &$2535\pm293$\\
\hline
\end{tabular}
\label{t2}
\end{table}
\subsection{Finite nuclei: neutron skin}

The EoS of infinite homogeneous nuclear matter calculated from the EDF can be
beneficently used to have estimates of some quantities relevant to
microscopic nuclear systems.  For example, with the calculated values
of $L_0$ and $K_{\rm sym,0}$, one can evaluate $\rho_A$, the
equivalent density of nuclei from the following equation, 
 \begin{equation}
e_{\rm sym}^s \simeq A^{1/3}[L_0\epsilon_A - \frac{1}{2}K_{\rm
sym,0}\epsilon^2_A],
 \end{equation}
where, $e_{\rm sym}^s$ is the surface symmetry energy coefficient. 
The equivalent density $\rho_A$ of a nucleus of mass $A$ is defined
as the density at which the symmetry coefficient $e_{\rm sym}(\rho_A)$ 
of nuclear matter
equals  $e_{sym}(A)$, the symmetry coefficient of the nucleus.
The 'experimental' value of $e_{\rm sym}^s$ is  
taken as 58.91$\pm $1.08 MeV \cite{Jiang12};
$\epsilon_A = (\rho_0 - \rho_A)/3\rho_0$. Fig. \ref{fig3} displays our calculated 
values of $\rho_A$ (shown as a shaded region) as a function of the
atomic mass number $A$. The blue triangles in the figure refer to the
value  of $\rho_A$ from Table I of Ref. \cite{Centelles09}, calculated
with different effective interactions for three nuclei, $A = 40, 116$
and $208$. The magenta line corresponds to the values calculated in
Ref. \cite{Liu10}. Our calculations with the indicated errors are seen
to have a good overlap with these results. With known $L_0$, an 
estimate of $r_{\rm skin}$, the neutron-skin thickness of heavy nuclei
can also be made. For that we make use of
the $L_0 - r_{\rm skin}$ correlation method as elucidated
in Refs. \cite{Centelles09,Agrawal13} for Skyrme interactions. As an
example, with our value of $L_0$ we obtain $r_{\rm skin}\simeq 0.21\pm
0.02$ fm for$^{208}$Pb. Some deliberations at this stage on the neutron
skin $ r_{\rm skin}$ of $^{208}$Pb may be meaningful. Recent PREX
experiment \cite{Abrahamyan12} reports a large central value of 0.33
fm for $r_{\rm skin}$ of $^{208}$Pb with very large error bars. This
contradicts nearly all the calculated results of $r_{\rm skin}$, which are
comparatively much smaller.  Fattoyev and Piekarewicz \cite{Fattoyev13}
devised a relativistic EoS that can accommodate such a large neutron
skin, but then $e_{sym}(\rho_0)$ and $L_0$ become uncomfortably high.
The larger the value of $r_{\rm skin}$, the larger becomes the value
of $L_0$.  It is known that the larger is then the neutron star  radius
\cite{Horowitz01a}. A large neutron star radius seems to be incompatible
with astrophysical data \cite{Lattimer14,Guillot13}, a very large value
for the neutron skin of $^{208}$Pb is thus doubtful. Calculations by
Brown \cite{Brown13} tend to disfavor a large neutron skin of $^{208}$Pb.
Nuclear ground state data for closed shell nuclei were fitted with a
set of Skyrme interactions with constraints of fixed $r_{\rm skin}$.
The average deviation for binding energies was found to be similar for
$r_{\rm skin}$ =0.16 and 0.20 fm, but increased by 0.1 to 0.3 MeV for
$r_{\rm skin}$ =0.24 fm.  A very recent experimental determination of
the neutron skin thickness from coherent pion production \cite{Tarbert14}
adds new dimension to this issue, the extracted value of $r_{\rm skin}$
for $^{208}$Pb is $r_{\rm skin} = 0.15 \pm 0.03$ fm.

\subsection{Supranormal densities : neutron stars }

Having come this far, we try to assess our EoS with reference to
that extracted  from experimental data at supranormal densities. This
is done in Fig.  \ref{fig4}. The upper panel displays the EoS (pressure
density relation) of symmetric nuclear matter (SNM). The shaded red
and yellow regions show the 'experimental' EoS for SNM synthesized
from collective flow data \cite{Danielewicz02} and from data for kaon
production \cite{Fuchs06,Fantina14}, respectively; the blue shaded
region shows ours. The EoS of pure neutron matter (PNM) has an additional
repulsive component coming from the density dependence of symmetry energy.
This part of EoS is laced with uncertainty, it is model dependent. The
lower panel shows the EoS of PNM. The shaded green region is the EoS of
PNM where the density dependence of symmetry energy is modeled as soft;
the red shaded region is the one where the said density dependence is
modeled as stiff \cite{Prakash88}.  The  blue shaded region displays the
results obtained in the present work; it has an excellent overlap with the
'experimental' EoS.  Possible phase transitions to exotic phases like
hyperons, kaons etc. at high densities  softens the EoS somewhat,
this is not taken into account in the present description.

For completeness, to  gauge the applicability of the EoS to higher
densities, we calculated the lower limit of the maximum mass of
the neutron star $(M_{max}^{NS})$ with this EoS solving the general
relativistic Tolman-Oppenheimer-Volkoff equation \cite{Weinberg72}.
The EoS for the crust was taken from the Baym, Pethick, Sutherland model
\cite{Baym71}. The EoS for the core region was calculated under the
assumption of  a charge-neutral uniform plasma of neutrons, protons,
electrons and muons in $\beta $-equilibrium. The EoS is causal for
$0 \le \rho \le 8.3 \rho_0 $, the central density of the neutron
star in our calculation never reaches beyond $\rho \sim 6\rho_0 $.
Our value of $M_{max}^{NS}$ (=  2.19 $M_\odot $) is consistent
with the currently observed value of 1.97$\pm $0.04 $M_\odot $ for the pulsar PSR
J1614-2230 \cite{Demorest10} and also the value of  2.01 $\pm $0.04
$M_\odot $ for the pulsar PSR J0348+0432 \cite{Antoniadis13}.
The presence of exotic degrees of freedom like hyperons in
the core of the neutron star is known to pull down the value
of $M^{NS}_{max}$ substantially \cite{Schulze11}, however, in
the relativistic mean-field (RMF) model, it is also seen that by
increasing the strength of coupling of the  hyperon  to the vector
mesons, the effect of hyperons on $M^{NS}_{max}$ can be much reduced
\cite{Weissenborn12}.  Recently analyzing different models, Lattimer {\it
et. al} \cite{Lattimer13,Lattimer14} constrained the value of the radius
$R_{1.4}$ for a neutron star of mass 1.4 $M_\odot$ to 12.1 $\pm $1.1 km
with 90 $\% $ confidence level; our values of  $R_{1.4}$ is 11.95 $\pm
$ 0.75 km.  Determination of neutron star radius is, however, not free
from uncertainty. Assuming that the neutron star core is best described
by 'normal matter' EoS, Guillot et. al \cite{Guillot13} find, again in
a 90 $\% $ confidence level, that for astrophysically relevant masses
($M_{NS} \ge 0.5 M_{\odot} $), the neutron star radius is quasi-constant,
$R_{NS}=9.1^{+1.3}_{-1.5} $km.

\begin{figure}[t]
\includegraphics[width=0.85\columnwidth,angle=0,clip=true]{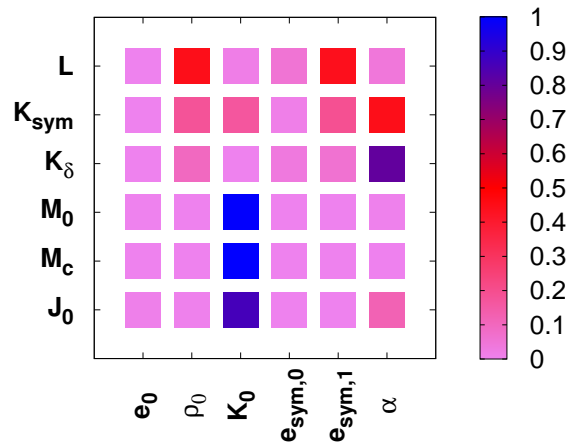}
\caption{\label{fig2} (Color online) The normalized squared errors, $\left (
\frac{\partial X}{\partial Y_i}\frac{\Delta Y_i}{\Delta X}\right )^2 (= x_i^2)$
are colour coded.  The quantities $X$ and $Y_i$ are given along the
ordinate and abscissa respectively. }
\end{figure}

\begin{table}[t]
\caption{\label{tab3}
The observables $X$ are listed in the first column, they are in units of MeV. 
The second column represents  $\Delta X$, the total uncertainty in $X$
from the input uncertainties $\Delta Y_i$ in the vector {\bf  Y} $\{e_0,\rho_0,
K_0,e_{\rm sym,0}, e_{\rm sym,1}, \alpha\}$.
The element $Y_2 (\equiv \rho_0)$ is in unit of fm$^{-3}$, $Y_6 (\equiv
\alpha)$ is dimensionless, all other elements in the vector {\bf  Y} are
in units of MeV. The units in the columns $\frac{\partial X}{\partial Y_i}
(i=1,..6)$ can then be obtained accordingly.}

\begin{tabular}{|c|r|r|r|r|r|r|r|}
\hline
$X$& $\Delta X$& $\frac {\partial X}{\partial Y_1}$& $\frac {\partial
X}{\partial Y_2}$& $\frac {\partial X}{\partial Y_3}$ &$\frac {\partial
X}{\partial Y_4}$& $\frac {\partial X}{\partial Y_5}$ &$\frac {\partial
X}{\partial Y_6}$\\ 
\hline
$L_0$&     13.5 &0.6   &   -1118 &     0.081  &  10.16  &  -11.09  &  -27.5    \\   
$K_{sym}$& 73.2& 12.2 &    -3889&     1.502  &  25.84  &  -40.07   &  -489    \\   
$K_\tau $&34.4 &-5.4 &    1339 &     0.024  &  -20.55  &  10.59   &  309.2    \\   
$M_0$&     293 &54   &    -447.5 &      14.6   &  0.00         &  0.00      &      -195   \\   
$ M_c$&    91 &-2.9 &    57.5   &      4.55 &  0.00         &  0.00        &      32.6 \\   
$Q_0$ &    56 &54   &    -447.5 &      2.6    &  0.00         &  0.00       &      -194.9  \\ 
\hline     
\end{tabular} 
\end{table}   

%\begin{table}[t]
%\caption{\label{tab4} 
%Partial errors $x_i^2 $ in $X$ due to uncertainty in $Y_i$; $\sum_i  x_i^2=1$.}
%\begin{tabular}{|c|r|r|r|r|r|r|}
%\hline
%$X$&  $x_1^2$&  $x_2^2$& $x_3^2$& $x_4^2$& $x_5^2$& $x_6^2$\\
%\hline
%$L_0$&        0.0000  &   0.4472&   0.0146 &  0.0555  &    0.4402 &    0.0423  \\
%$K_{sym}$&    0.0002  &   0.1807&   0.1685 &  0.0119  &    0.1919 &    0.4465  \\
%$K_\tau $&   0.0002  &   0.0969&   0.0001 &  0.0342  &    0.0606 &    0.8077  \\
%$M_0$&        0.0003  &   0.0001&   0.9950 &  0.0000             &    0.0000     &    0.0044 \\
%$ M_c$&       0.0000 &    0.0000 &   0.9986 &  0.0000              &    0.0000      &    0.0012 \\
%$Q_0$ &       0.0093 &   0.00410 &   0.8650 &  0.0000                &    0.0000                 &
%0.1215 \\
%\hline 
%\end{tabular}
%\end{table}

%\clearpage
\begin{figure}
\includegraphics[width=0.85\columnwidth,angle=0,clip=true]{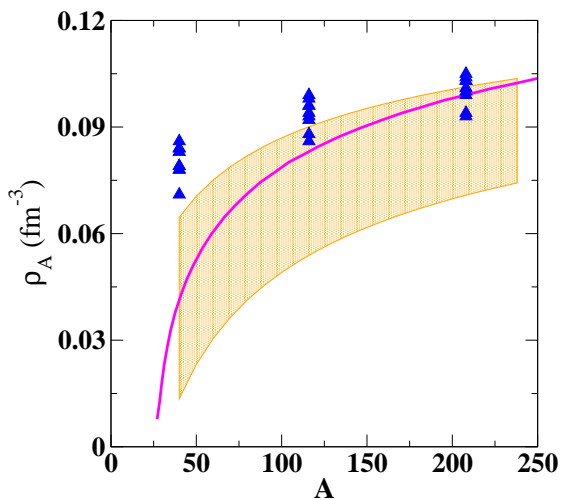}
\caption{(Color online) {\label{fig3}  Our results for
the equivalent density shown as a shaded
region as a function of mass number. The blue triangles are from
Ref. \cite{Centelles09}, the magenta line is from Ref. \cite{Liu10}.  }}
\end{figure}

\begin{figure}
\includegraphics[width=0.85\columnwidth,angle=0,clip=true]{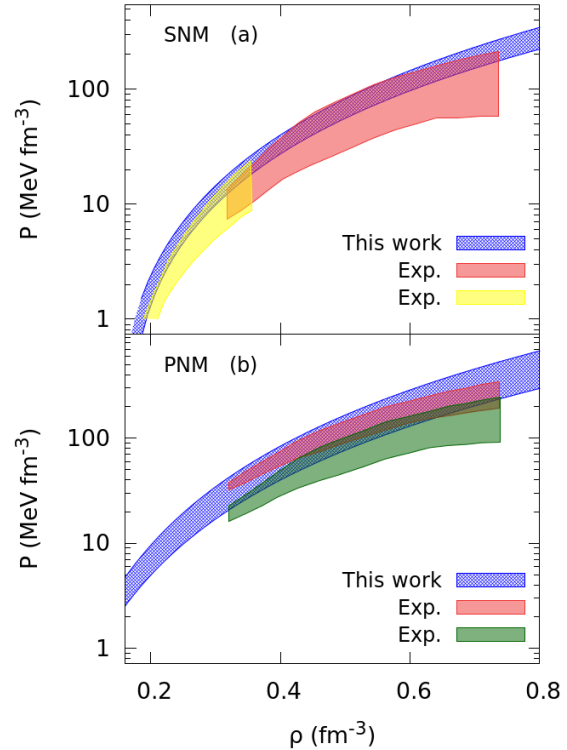}
\caption{(Color online) \label{fig4}  The EoS for symmetric nuclear matter
(upper panel) and for pure neutron matter (lower panel).   The red,
yellow and  green  shaded regions represent the experimental data taken
from Ref. \cite{Danielewicz02,Fuchs06,Fantina14,Prakash88}. The
blue shaded  regions are the EoS obtained in this work. See text for
details.}\end{figure}

\section{Conclusions}

To sum up, from  consensus 'empirical' inputs for values of some of the
key nuclear parameters at saturation and sub-saturation densities,
we have constructed a Skyrme-type energy density functional for
homogeneous nuclear matter.  This is then employed to  
understand the density dependence of the nuclear
symmetry energy and incompressibility  and to predict values
for the important nuclear parameters like the symmetry slope parameter
$L_0$, the symmetry incompressibility parameter $K_{\tau }$, the
incompressibility slope parameter $M(\rho )$.  Separate estimates of
these quantities have been given from different perspectives; sizeable
uncertainties remain there.  The structural edifice for the energy
density functional built on a few known input bulk parameters gives a
coherence in the evaluated values of the observables; their uncertainties
can be constrained better provided the input bulk entities are known with
better precision. The general agreement of our EoS with the 'experimental'
one at supranormal densities is interestingly  striking. 
The near concordance of our calculated lower bound of the maximum
mass of neutron star with the experimental observation of a neutron
star of mass $M^{NS}_{max} \sim 2M_\odot $ is also very noticeable.
Inclusion of exotic degrees of freedom in the interior of the star,
however, softens the EoS and lowers the value of $M^{NS}_{max}$
though and this needs further investigation.

%\Acknowledgement
\begin{acknowledgments}
JND acknowledges support from the Department of Science and Technology,
Government of India.  G.C. would like to thankfully acknowledge the
nice hospitality extended to him during his visit to SINP, when this
work started.  The authors gratefully acknowledge the assistance of
Tanuja Agrawal in the preparation of the manuscript.
\end{acknowledgments}

%\bibliography{plb_ref}

\end{document}